\documentclass[aps,prl,superscriptaddress,amsmath,amssymb,twocolumn,nofootinbib]{revtex4}

\usepackage[utf8]{inputenc}
\usepackage{color}
\usepackage{graphicx}
\usepackage{dashbox}
\usepackage{physics}

\begin{document}

\thispagestyle{empty}

\title{GREA and Dark Energy: A holographic correspondence}

\author{Juan Garc\'ia-Bellido}\email[]{juan.garciabellido@uam.es}
\affiliation{Instituto de F\'isica Te\'orica UAM-CSIC, Universidad Auton\'oma de Madrid, Cantoblanco 28049 Madrid, Spain}

\date{November 24, 2025}

\preprint{IFT-UAM/CSIC-24-149}

\begin{abstract}
The nature of the cosmological constant is a mystery. We don't understand its quantum origin but we associate it with the actual acceleration of the universe because it is the simplest description we had until recently of the present cosmological observations. However, this may change with the next generation of experiments. If we can convince ourselves that the cosmic acceleration is not due to a constant, this would open up new fascinating avenues. By exploring the simplest cosmological model in the bulk, that of an empty and flat space with a cosmological constant $\Lambda$, we find that its holographic correspondence makes sense as a theory of fundamental quantum degrees of freedom at the boundary. Moreover, we find that an observer in the bulk, making long-range gravitational observations, cannot distinguish the acceleration induced by the cosmological constant $\Lambda$ from that induced by the thermodynamic properties of the boundary, the de Sitter horizon, strictly at the level of the background cosmology. By including matter in the bulk we extend this holographic correspondence to GREA, where the quantum d.o.f. associated with the evolving boundary of the causal horizon induces an entropic acceleration that varies in time. Upcoming surveys such as DESI, Euclid, and the Vera Rubin Observatory (LSST) will test this framework through the growth of large-scale structures in the late universe, where GREA and $\Lambda$CDM differ quantitatively.
\end{abstract}
\maketitle

\section{I. Introduction}

General Relativistic Entropic Acceleration (GREA) theory~\cite{Espinosa-Portales:2021cac} gives a covariant formalism for out of equilibrium dynamics in the context of general relativity. The main feature of GREA is the explicit breaking of the time reversal invariance when there is entropy production. This drives an entropic force that behaves effectively like a bulk viscosity with a negative effective pressure. When the entropy production is associated with horizons one can describe their thermodynamical effects as a GHY boundary term~\cite{Gibbons:1976ue,York:1972sj}. The natural consequence of this formalism is that the source for space-time curvature is the Helmholtz free energy, $F = U - TS$, and not just matter and radiation. In the cosmological context, this theory predicts that
the present acceleration of the universe should arise from the growth of entropy associated with cosmic~\cite{Garcia-Bellido:2021idr} and black hole horizons~\cite{Garcia-Bellido:2024tip}, without the need to introduce a cosmological constant. 

The present paper makes a new contribution with respect to the prior GREA literature~\cite{Espinosa-Portales:2021cac,Garcia-Bellido:2021idr,Garcia-Bellido:2024tip,Garcia-Bellido:2024qau}: whereas those works used the substitution of the cosmological constant by the GHY boundary term as a tool to reproduce the observed acceleration, here we ask whether the holographic correspondence between the boundary term and the bulk cosmological constant, Eq.~(\ref{eq:holo}), is \emph{fundamental}. We show that the Hawking-Gibbons thermodynamics of the de~Sitter boundary provides an explicit quantum thermodynamic interpretation of~$\Lambda$: the quantum vacuum energy in the bulk is holographically correspondent to the quantum entropy of unknown degrees of freedom on the boundary. Unlike AdS/CFT, where a complete dual theory and an explicit dictionary exist, the present correspondence does not yet specify the boundary degrees of freedom or provide a calculable boundary theory --- this remains an important open problem.

\section{II. The Einstein-Hilbert action in GREA}

We give here a short summary of the covariant formalism developed in Ref.~\cite{Espinosa-Portales:2021cac} that describes the extended Einstein field equations in the presence of out-of-equilibrium phenomenon associated with the growth of horizons.
The gravitational action of the Standard Model of Cosmology ($\Lambda$CDM) is given by
$${\cal S} = \int_{\cal M}\!d^4x
\sqrt{-g}\,\Big[\frac{1}{2\kappa}R+{\cal L}_m\Big] + 
\int_{\cal M}\!d^4x\sqrt{-g}\,\frac{(-\Lambda)}{\kappa}\,,$$
where ${\cal L}_m$ is the matter Lagrangian, $\kappa = 8\pi G$ and $\Lambda$ is the cosmological constant. 
In the context of GREA we substitute the bulk term of the cosmological constant for the GHY boundary term of the horizon~\cite{Garcia-Bellido:2021idr},
$${\cal S} = \int_{\cal M}\!d^4x\sqrt{-g}\,
\Big[\frac{1}{2\kappa}R+{\cal L}_m\Big]  + 
\int_{\cal H}\!d^3x\sqrt{h}\,\frac{K}{\kappa}\,,$$
where $K$ is the extrinsic curvature and $h$ the 3D metric of the cosmological horizon, the boundary ${\cal H}=\partial{\cal M}$.

The question that arises is whether this holographic correspondence is fundamental or not,
\begin{equation}\label{eq:holo}
\int_{\partial{\cal M}}\!\!\!d^3x\sqrt{h}\,K =
\int_{\cal M}\!d^4x\sqrt{-g}\,(-\Lambda)\,.
\end{equation}
At face value, it is the statement that, from the point of view of the dynamics of space-time, one cannot distinguish the effect of a {\em constant} ``matter" component in the bulk with $\kappa{\cal L}_m = - \Lambda$ from that of the de Sitter boundary, the future conformal boundary of space, ${\cal H} = \partial{\cal M}$. Note that such a matter Lagrangian gives rise to a {\em classical} constant energy density in the bulk, with $\Lambda = \kappa\rho_V$. The matter contribution to the Hamiltonian constraint, $3H^2=\kappa\rho_V$, arises from variations of the matter action with respect to the lapse function (and afterwards setting $N(t)=c=1$), for a flat and empty FLRW universe\footnote{Homogeneity and isotropy seems to be a fundamental assumption, but spatial flatness is not required, it only simplifies the expressions. Note that $N(t)$ is kept explicit in the action in order to perform the variation $\delta{\cal S}/\delta N = 0$ that yields the Hamiltonian constraint; it is set to unity in the resulting equations of motion, as is standard in the ADM formalism.}
\begin{equation}
	\label{eq:matter}
	{\cal S}_m = \int\!d^4x\sqrt{-g}\,{\cal L}_m = - \int dt
    \,N(t)\,\frac{4\pi}{3}\,r_H^3\rho_V\,,
\end{equation}
where $r_H$ is the physical radius of the dS horizon, which is constant and can be computed in terms of the local rate of expansion,
$$r_H = H^{-1} = \sqrt{\frac{3}{\Lambda}}\,,$$
which only depends on the {\it a priori constant} $\Lambda$.

\subsection{Interpretation of Hawking-Gibbons action}

What is surprising is that the quantum origin of the cosmological constant is not manifest in the classical $\Lambda$CDM model. In the action above, $\Lambda$ is a constant and is  indistinguishable from that of the vacuum energy of quantum fields {\em in the bulk}. We don't have a way to compute this vacuum energy exactly and the estimates that are performed fail miserably by many orders of magnitude~\cite{Weinberg:1988cp}. Perhaps we are missing some fundamental ingredient. 

I propose here that the holographic correspondence (\ref{eq:holo}) may give us a clue to the quantum interpretation of the cosmological constant. Let us consider a comoving sphere around the origin of coordinates $r=0$ with unit normal vector $n_r = a^{-1}\partial_r$. Then the trace of the extrinsic curvature is given by 
$\sqrt{h} K = 2N(t)\,r\,a\,\sin\theta$, and the GHY term for the dS horizon becomes~\cite{Garcia-Bellido:2021idr}
\begin{equation}\label{eq:GHY}
       \!{\cal S}_{\rm GHY} = -\frac{1}{2G}\int\!dt\,N(t)\,H\,r_H^2 = -\!\int\!dt\,N(t)\,T_H\,S_H 
\end{equation}
where $T_H$ is the temperature and $S_H$ is the total entropy associated with the dS horizon~\cite{Gibbons:1976ue},
\begin{equation}\label{eq:STH}
    k_{_{\rm B}}T_H = \frac{\hbar\,H}{2\pi}\,,
    \hspace{1cm}
    S_H = \frac{k_{_{\rm B}}}{\hbar}\frac{4\pi\,r_H^2}{4G}
    \,.
\end{equation}
Note that although the origin of the Hawking-Gibbons temperature (and corresponding entropy) is quantum mechanical, gravitational and thermodynamical, the contribution to the Hamiltonian constraint is classical, since fundamental constants ($\hbar, k_{_{\rm B}}$) cancel.

There is however a holographic dual description of the energy density that drives the acceleration due to a cosmological constant in terms of the Hawking-Gibbons thermodynamics in the boundary,
\begin{equation}\label{eq:STH}
    T_H\,S_H = \frac{4\pi}{\kappa H} = \frac{4\pi}{3}r_H^3\rho_V = \frac{4\pi}{\kappa}\sqrt{\frac{3}{\Lambda}}\,.
\end{equation}
Comparing Eqs.~(\ref{eq:matter}) and (\ref{eq:GHY}), together with (\ref{eq:STH}), we see that the GREA force~\cite{Espinosa-Portales:2021cac} associated with the boundary of dS is indistinguishable from the acceleration induced by the cosmological constant in the bulk, strictly at the level of the Hamiltonian constraint (background cosmology). In the pure de Sitter case there is no matter, so $U=0$ and the Helmholtz free energy reduces to $F = -T_H S_H$, which is precisely the term computed in Eq.~(\ref{eq:STH}); matter and radiation contribute to $U$ and modify the dynamics at the perturbative level~\cite{Garcia-Bellido:2024qau}. In other words, the quantum vacuum energy in the bulk is holographic correspondent to the quantum entropy of {\em unknown} quantum degrees of freedom on the boundary of de Sitter space.

Note that the entropy of de Sitter is constant and never\-theless induces a GREA on galaxies in the bulk. Both arise from variations w.r.t. the lapse function $N$, which gives the Hamiltonian constraint, ${\cal H} = 0$, associated with time-reparametrization invariance.
From the point of view of general covariance, a constant surface term is indistinguishable from a cosmological constant in the bulk. In other words, one cannot distinguish the acceleration induced by vacuum energy from that of the entropic force associated with the GHY boundary term. This is probably the simplest realization of the holographic correspondence~\cite{tHooft:1993dmi,Susskind:1994vu}. Unfortunately, this does not give us a clue as to the origin of the quantum degrees of freedom at the boundary that are responsible for the Bekenstein-Hawking entropy of de Sitter.

\section{Including matter and radiation: A moving boundary of space}

\begin{figure}
    \centering
    \includegraphics[width=0.99\linewidth]{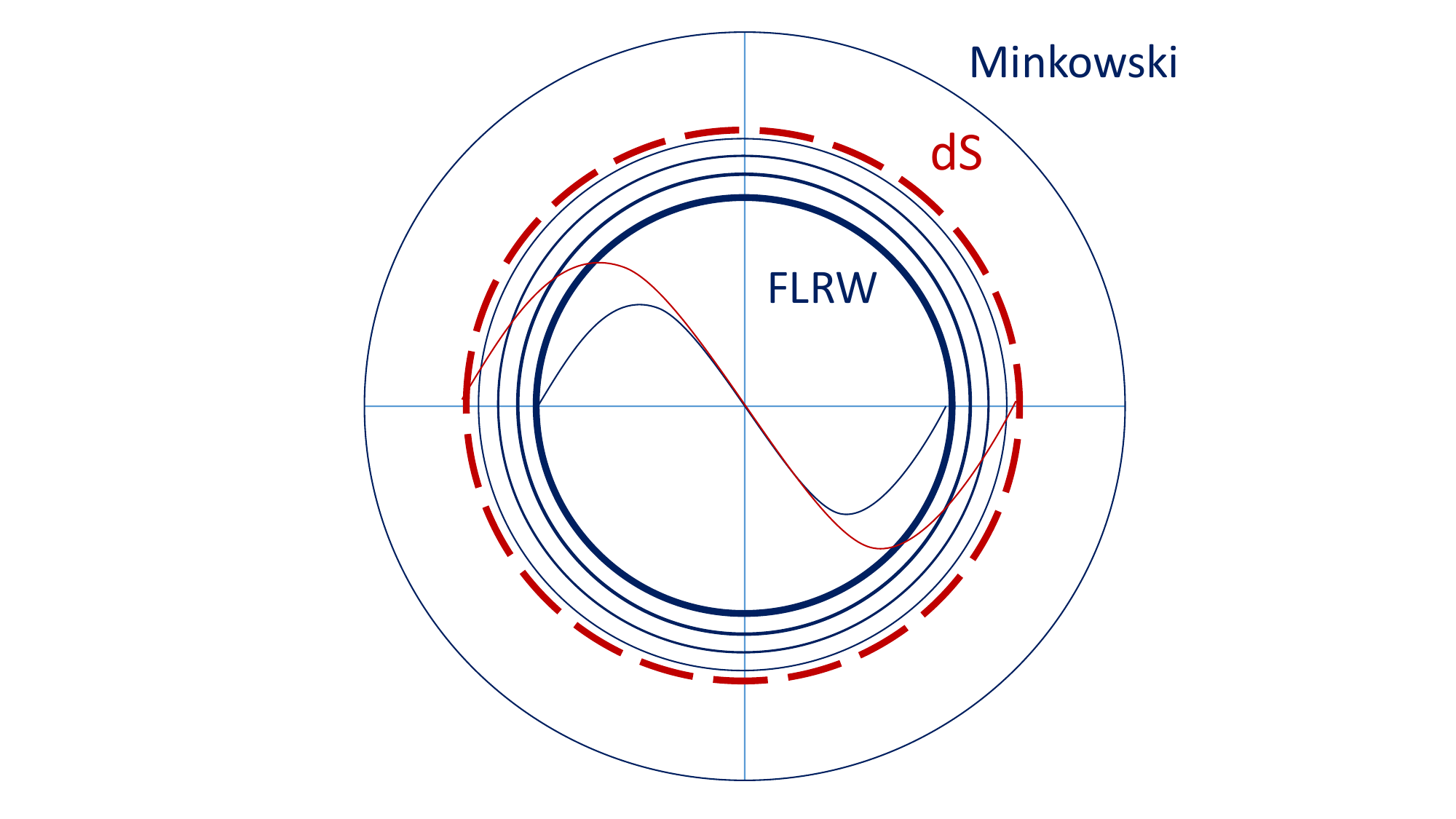}
    \caption{The asymptotic de Sitter horizon (red) and the evolving FLRW horizon (blue) in the presence of matter and radiation. Also shown are the quantum field modes with the largest wavelengths that fit inside the horizon. In the absence of a cosmological constant, the asymptotic space-time is empty and flat Minkowski space-time.}
    \label{fig:Horizons}
\end{figure}

The de Sitter horizon is the asymptotic boundary of space when all matter has been diluted away and we are left with an empty universe with a cosmological constant. However, in the presence of matter and radiation, the causal horizon of our universe is not constant, it evolves with time, see Fig.~\ref{fig:Horizons}. The causal horizon grows as those components dilute away. Such a boundary must also give rise to a thermodynamical description {\em a la} Gibbons-Hawking, even though in this case out-of-equilibrium, since the entropy in such a boundary is not constant but grows, slowly at the beginning but then faster and faster~\cite{Garcia-Bellido:2024qau} as matter dilutes. This growth of entropy is not a postulate but a geometric consequence of the expansion: the Bekenstein-Hawking entropy $S_H \propto r_H^2$ grows monotonically as the comoving causal horizon $r_H$ expands with the dilution of matter and radiation; the covariant derivation of the resulting entropic force is given in Ref.~\cite{Espinosa-Portales:2021cac}.

The main difference of GREA with respect to $\Lambda$CDM is the breaking of time-reversal invariance at the level of the action by the entropic force term that arises due to the growth of the cosmic horizon and thus to the growth of entropy. The explicit breaking of time-reversal invariance at the level of the action by the entropy-production term suggests a natural cosmological arrow of time, whose quantitative implications are left for future work. Most of the evolution of the universe has been quasi-adiabatic since these entropic forces acted only during inflation, reheating after inflation and at the present time, when the universe is old enough and the cosmic horizon associated with radiation and matter has grown sufficiently large for its entropy growth to dominate over the attraction of matter and induce the present acceleration~\cite{Garcia-Bellido:2021idr}. Eventually, even this period of acceleration will end as the entropic term gets diluted and we will end in an empty Minkowski space-time~\cite{Garcia-Bellido:2024qau}. If there were a non-vanishing cosmological constant, the causal cosmological horizon induced by matter will asymptotically approach the future de Sitter horizon and instead of Minkowski the universe would end up in de Sitter space-time.


\section{Acknowledgements}

The author acknowledges support from the Spanish Research Project PID2024-159420NB-C43 [MICINN-FEDER], and the Centro de Excelencia Severo Ochoa Program CEX2020-001007-S at IFT. 

\bibliography{main}

@article{Weinberg:1988cp,
    author = "Weinberg, Steven",
    editor = "Hsu, Jong-Ping and Fine, D.",
    title = "{The Cosmological Constant Problem}",
    reportNumber = "UTTG-12-88",
    doi = "10.1103/RevModPhys.61.1",
    journal = "Rev. Mod. Phys.",
    volume = "61",
    pages = "1--23",
    year = "1989"
}

@article{Espinosa-Portales:2021cac,
    author = "Espinosa-Portal\'es, Llorenc and Garc\'ia-Bellido, Juan",
    title = "{Covariant formulation of non-equilibrium thermodynamics in General Relativity}",
    eprint = "2106.16012",
    archivePrefix = "arXiv",
    primaryClass = "gr-qc",
    reportNumber = "IFT-UAM/CSIC-21-74",
    doi = "10.1016/j.dark.2021.100893",
    journal = "Phys. Dark Univ.",
    volume = "34",
    pages = "100893",
    year = "2021"
}

@article{Garcia-Bellido:2021idr,
    author = "Garc\'ia-Bellido, Juan and Espinosa-Portal\'es, Llorenc",
    title = "{Cosmic acceleration from first principles}",
    eprint = "2106.16014",
    archivePrefix = "arXiv",
    primaryClass = "gr-qc",
    reportNumber = "IFT-UAM/CSIC-21-75",
    doi = "10.1016/j.dark.2021.100892",
    journal = "Phys. Dark Univ.",
    volume = "34",
    pages = "100892",
    year = "2021"
}

@article{Garcia-Bellido:2024tip,
    author = "Garc\'\i{}a-Bellido, Juan",
    title = "{Cosmic entropic acceleration from supermassive black hole growth}",
    eprint = "2306.10593",
    archivePrefix = "arXiv",
    primaryClass = "gr-qc",
    reportNumber = "IFT-UAM/CSIC-23-72",
    doi = "10.1016/j.dark.2024.101491",
    journal = "Phys. Dark Univ.",
    volume = "44",
    pages = "101491",
    year = "2024"
}

@article{York:1972sj,
    author = "York, Jr., James W.",
    title = "{Role of conformal three geometry in the dynamics of gravitation}",
    doi = "10.1103/PhysRevLett.28.1082",
    journal = "Phys. Rev. Lett.",
    volume = "28",
    pages = "1082--1085",
    year = "1972"
}

@article{Gibbons:1976ue,
    author = "Gibbons, G. W. and Hawking, S. W.",
    title = "{Action Integrals and Partition Functions in Quantum Gravity}",
    reportNumber = "PRINT-76-0995 (CAMBRIDGE)",
    doi = "10.1103/PhysRevD.15.2752",
    journal = "Phys. Rev. D",
    volume = "15",
    pages = "2752--2756",
    year = "1977"
}

@article{tHooft:1993dmi,
    author = "'t Hooft, Gerard",
    title = "{Dimensional reduction in quantum gravity}",
    eprint = "gr-qc/9310026",
    archivePrefix = "arXiv",
    reportNumber = "THU-93-26",
    journal = "Conf. Proc. C",
    volume = "930308",
    pages = "284--296",
    year = "1993"
}

@article{Susskind:1994vu,
    author = "Susskind, Leonard",
    title = "{The World as a hologram}",
    eprint = "hep-th/9409089",
    archivePrefix = "arXiv",
    reportNumber = "SU-ITP-94-33",
    doi = "10.1063/1.531249",
    journal = "J. Math. Phys.",
    volume = "36",
    pages = "6377--6396",
    year = "1995"
}

@article{Garcia-Bellido:2024qau,
    author = "Garc\'ia-Bellido, Juan",
    title = "{Dark Energy predictions from GREA: Background and linear perturbation theory}",
    eprint = "2405.02895",
    archivePrefix = "arXiv",
    primaryClass = "astro-ph.CO",
    reportNumber = "IFT-UAM/CSIC-24-69",
    doi = "10.1016/j.dark.2024.101533",
    journal = "Phys. Dark Univ.",
    volume = "45",
    pages = "101533",
    year = "2024"
}

\end{document}